\documentclass[oneside,a4paper,reqno]{amsart}
\usepackage{amsmath}
\usepackage{amssymb}
\usepackage[mathscr]{eucal}
\usepackage{graphics}

\newcommand{\bmat}[3]{\bigl \langle   #1 \bigr| \, #2\, \bigl|     #3  
\bigr \rangle}

\newcommand{\vmat}[2]{\langle #1, \, #2\rangle}
\newcommand{\bvmat}[2]{\bigl< #1, \, #2 \bigr>}

\newcommand{\overlap}[2]{	\left \langle    #1 \vphantom{#2 } \,
                        \right| \left.   #2 \vphantom{#1}
                        \right \rangle	}
\newcommand{\boverlap}[2]{ 
	\bigl \langle #1 \, \bigr| \bigl. #2 \bigr\rangle	}

\newcommand{\weyl}[1]{ #1_{\mathrm{W}}}
\newcommand{\mbrack}[2]{\left\{#1,#2\right\}_{\mathrm{W}}}
\newcommand{\hus}[1]{#1_{ \mathrm{H} 
    \vphantom{ \overline{ \mathrm{H} } }}}
\newcommand{\anti}[1]{#1_{\overline{\mathrm{H}}}}
\newcommand{\hfunc}{Q}
\newcommand{\hbrack}[2]{\left\{#1,#2\right\}_{\mathrm{H}}}
\newcommand{\dplus}{\partial_{+}}
\newcommand{\dminus}{\partial_{-}}
\newcommand{\dplmin}{\partial_{\pm}}
\DeclareMathOperator{\Tr}{Tr}
\DeclareMathOperator{\Imag}{Im}
\theoremstyle{plain}
\newtheorem{theorem}{Theorem}
\newtheorem{lemma}[theorem]{Lemma}
\begin{document}\begin{titlepage}
\begin{center}
\bfseries
HUSIMI TRANSFORM OF AN OPERATOR PRODUCT
\end{center}
\vspace{1 cm}
\begin{center}
D M APPLEBY
\end{center}
\begin{center}
Department of Physics, Queen Mary and
		Westfield College,  Mile End Rd, London E1 4NS, UK
 \end{center}
\vspace{0.5 cm}
\begin{center}
  (E-mail:  D.M.Appleby@qmw.ac.uk)
\end{center}
\vspace{0.75 cm}
\vspace{1.25 cm}
\begin{center}
\textbf{Abstract}\\
\vspace{0.35 cm}
\parbox{10.5 cm }{ It is shown that the series derived by
Mizrahi, giving the Husimi transform (or covariant symbol) of
an operator product, is absolutely convergent for a large class of
operators.  In  particular, the generalized Liouville
equation, describing the time evolution of the Husimi
function, is absolutely convergent for a large class of
Hamiltonians.  By contrast, the series derived by
Groenewold, giving the Weyl transform of an operator product,
is often only asymptotic, or even undefined.  The result is
used to derive an alternative way of expressing expectation
values   in terms of the Husimi function.   The advantage of this
formula is that it applies in many of the cases where the
anti-Husimi transform (or contravariant symbol) is so highly
singular that it fails to exist as a tempered distribution.
                      }
\end{center}
\vspace{1 cm}
\end{titlepage}
\section{Introduction}
\label{sec:  intro}
A particularly useful and illuminating way of studying the
classical limit is to formulate quantum mechanics in terms of
phase space distributions~\cite{Hill,Lee,Leon}.  The
advantage of such a formulation as compared with the standard Hilbert
space formulation  is that it puts quantum mechanics into
a form which is similar to the probabilistic phase space
formulation of classical mechanics.  At least from a formal,
mathematical point of view it thus allows one to regard
quantum mechanics as a kind of generalized version of classical
mechanics.

There are, of course, many different phase space formulations of
quantum mechanics.  The one which was discovered first is the
formulation based on the Wigner
function~\cite{Hill,Lee,Leon,Wigner}. In the case of a system
having one degree of freedom with position
$\hat{x}$, momentum
$\hat{p}$ and density matrix
$\hat{\rho}$ the Wigner function is defined by
\begin{equation*}
  W(x,p) = \frac{1}{2 \pi}\int dy \, e^{i p y}
\bmat{x-\tfrac{y}{2}}{\hat{\rho}}{x+\tfrac{y}{2}}
\end{equation*}
(in units chosen such that $\hbar=1$).
The Wigner function continues to find many important applications
(in quantum tomography~\cite{Leon}, for example).
However, if the aim is specifically to represent quantum
mechanics in a manner which resembles classical mechanics as
closely as possible, then the Wigner function suffers from the
serious disadvantage that it is not strictly non-negative (except
in special cases~\cite{PosWig})---which makes the analogy with the
classical phase space probability distribution somewhat strained.

There has accordingly been some interest in the problem of
constructing alternative distributions, which are strictly non-negative,
and which can be interpreted as probability density functions.   There
are, in fact, infinitely many such
functions~\cite{Davies,Cart,Wod1,Dav1,Halli,Wun}.   The one which
was discovered first, and which is the focus of this paper, is the Husimi,
or
$Q$-function~\cite{Hill,Lee,Leon,Hus,Kano,Glaub,Miz}, 
which is
obtained from the Wigner function by smearing it with a Gaussian
convolution:
\begin{equation*}
  \hfunc(x,p)
=  \frac{1}{\pi} \int dx' dp' \,
  \exp\left[ -(x-x')^2 - (p-p')^2 \right]
   W(x',p')
\end{equation*}
(in units such that $\hbar=1$, and where we assume that 
$x$, $p$ have been made dimensionless by choosing a suitable
length scale $\lambda$, and making the replacements
$x\rightarrow x/\lambda$, 
$p\rightarrow \lambda p$).
It should be emphasised that it is not simply that the Husimi function
has the mathematical significance of a probability density function.  It
also has this significance physically.  It has been shown that the
Husimi function is the probability distribution describing the outcome
of a joint measurement of position and momentum in a number of
particular cases~\cite{Leon,Wod1,Art,Raymer,Leon2}.  More
generally it can be shown~\cite{Ali,self1} that the Husimi function
has a universal significance:  namely, it is the probability density
function describing the outcome of \emph{any} retrodictively optimal
joint measurement process.  In Appleby~\cite{self1} it is argued that
this means that the Husimi function may be regarded as the canonical
quantum mechanical phase space probability distribution, which plays
the same role in relation to joint measurements of $x$ and $p$ as does
the function
$\left| \overlap{x}{\psi}\right|^2$ in relation to single measurements
of $x$ only.

If one wants to construct a systematic procedure for investigating the
transition from quantal to classical it is not enough simply to find an
analogue for the classical phase space probability distribution.  One
also needs an analogue of the classical Liouville equation, giving the
time evolution of the probability distribution.  In the formulation based
on the Husimi function this is accomplished by means of Mizrahi's
formula~\cite{Miz}, giving the Husimi transform of an operator
product (also see Lee~\cite{Lee}, Cohen~\cite{Cohen},
Prugove\v{c}ki~\cite{Prugo} and 
O'Connell and Wigner~\cite{Conn}).

Let $\weyl{A}$ denote the Weyl
transform of the  operator $\hat{A}$, defined 
by~\cite{Hill,Lee,Weyl}
\begin{equation*}
  \weyl{A}(x,p) = \int dy \, e^{i p y}
\bmat{x-\tfrac{y}{2}}{\hat{A}}{x+\tfrac{y}{2}}
\end{equation*}
The Husimi transform (or covariant symbol) 
$\hus{A}$ is then given by
\begin{equation}
  \hus{A}(x,p)
=  \frac{1}{\pi} \int dx' dp' \,
  \exp\left[ -(x-x')^2 - (p-p')^2 \right]
   \weyl{A}(x',p')
\label{eq:  HusDef}
\end{equation}
Mizrahi~\cite{Miz} has derived the following formula for the Husimi 
transform of the product of two operators $\hat{A}$,
$\hat{B}$:
\begin{equation}
  \hus{ (\hat{A}\hat{ B})}
=  \hus{A} e^{\overleftarrow{\dplus} \overrightarrow{\dminus}}
   \hus{B}
\label{eq:  HusProd}
\end{equation}
where 
$ \dplmin =2^{-1/2}(\partial_{x} \mp i \partial_{p})$.  Using this
formula, and the fact that the Husimi function is just the Husimi
transform of the density matrix scaled by a factor $1/(2 \pi)$,
 it is 
straightforward to derive the following generalization of the Liouville
equation:
\begin{equation}
  \frac{\partial}{\partial t} \hfunc
=  \hbrack{\hus{H}}{ \hfunc}
\label{eq:  HusLiouGen}
\end{equation}
where $\hus{H}$ is the Husimi transform of the
Hamiltonian, and $\hbrack{\hus{H}}{
\hfunc}$ is the generalized Poisson bracket
\begin{equation}
  \hbrack{\hus{H}}{\hfunc}
=  \sum_{n=0}^{\infty}
     \frac{2}{n!} \Imag \left( \dplus^{n}\hus{H} \,
     \dminus^{n}\hfunc\right)
\label{eq:  HusBracket}
\end{equation}
The first term in the sum on the right-hand side is just the
ordinary Poisson bracket.  The remaining terms represent quantum
mechanical corrections.

It is not apparent from Mizrahi's derivation, whether these expressions
are exact, or whether they are only asymptotic.
In Section~\ref{sec:  ProdForm} we will show that there is a large class of
operators for which the series in Eq.~(\ref{eq:  HusProd})  
[and consequently the series in
Eq.~(\ref{eq:  HusBracket})]
is absolutely convergent .  This property is closely connected with
the complex analytic properties of the Husimi transform, as discussed by 
Mehta and Sudarshan~\cite{Mehta} and  Appleby~\cite{self2}.

The significance of the result proved in 
Section~\ref{sec:  ProdForm} is best appreciated if one compares
Eqs.~(\ref{eq:  HusProd}--\ref{eq:  HusBracket}) with the
corresponding formulae in the Wigner-Weyl 
formalism~\cite{Hill,Lee,Groen,Moyal,WeylProd}:
\begin{align}
  \weyl{(\hat{A} \hat{B})} 
& = \weyl{A} \exp\left[ \tfrac{i}{2}
      \left(\overleftarrow{\partial_{x}}
            \overrightarrow{\partial_{p}}-
             \overleftarrow{\partial_{p}}
            \overrightarrow{\partial_{x}}
       \right) \right]
  \weyl{B}
\label{eq:  WeylProduct}
\\
  \frac{\partial}{\partial t} W
& =  \mbrack{ \weyl{H}}{ W}
\label{eq:  LiouGen}
\\
\left\{ \weyl{H}, W\right\}_W
& = \sum_{n=0}^{\infty}
     \frac{(-1)^n}{(2n+1)! 2^{2 n}}
     H_W 
      \left(\overleftarrow{\partial_{x}}
            \overrightarrow{\partial_{p}}-
             \overleftarrow{\partial_{p}}
            \overrightarrow{\partial_{x}}
       \right)^{2 n+1}
    W
\label{eq:  MoyBrack}
\end{align}
where $\weyl{H}$ is the Weyl transform of the Hamiltonian,
and where $\mbrack{ \weyl{H}}{ W}$ denotes the Moyal
bracket~\cite{Moyal}.

It can be seen that Eqs~(\ref{eq:  HusProd}--\ref{eq:  HusBracket})
and Eqs.~(\ref{eq:  WeylProduct}--\ref{eq:  MoyBrack}) are formally
very similar.  However, this formal resemblance is somewhat
deceptive,  for it turns out that  the  two sets of equations have quite
different convergence properties.

The formula for the Weyl transform of an operator
product, Eq.~(\ref{eq:  WeylProduct}), is  exact if either
$\hat{A}$ or
$\hat{B}$ is a polynomial in $\hat{x}$ and $\hat{p}$ (in which
case the series terminates after a finite number of terms). 
More generally, if $\weyl{A}$, $\weyl{B}$ are $C^{\infty}$
functions satisfying appropriate conditions on their growth at
infinity, then it can be shown that the series is
asymptotic~\cite{Omnes}.  However, there are many operators of
physical interest for which the Weyl transform is only defined
in a distributional sense, and for operators such as this the
series can be highly singular.
Consider, for example, the parity operator $\hat{V}$, whose
action in the
$x$-representation is given by
\begin{equation*}
  \bmat{x}{\hat{V}}{\psi} = \boverlap{-x}{\psi}
\end{equation*}
We have
\begin{equation*}
  \weyl{V}(x,p)=\pi \delta(x)\delta(p)
\end{equation*}
Substituting this expression into  
Eqs.~(\ref{eq:  WeylProduct}) gives
\begin{equation*}
  \weyl{(\hat{V}^2)}(x,p)
=  \pi^2 \delta(x) \delta(p)
    \exp\left[ \tfrac{i}{2}
      \left(\overleftarrow{\partial_{x}}
            \overrightarrow{\partial_{p}}-
             \overleftarrow{\partial_{p}}
            \overrightarrow{\partial_{x}}
       \right) \right]
    \delta(x) \delta(p)
\end{equation*}
The left-hand side of this equation $=1$; whereas
the expression on the right-hand side is an infinite sum each
individual term of which is ill-defined (being a product of
distributions concentrated at the origin).

Mizrahi's~\cite{Miz} derivation of the formula for the Husimi
transform of an operator product, Eq.~(\ref{eq:  HusProd}), depends
on the same kind of  formal manipulation that is used in
Groenewold's~\cite{Groen} derivation of  Eq.~(\ref{eq: 
WeylProduct}), and so it might be supposed that the validity of
the formula is similarly restricted.  However, it turns out that the
sum in Eq.~(\ref{eq:  HusProd}) is actually much better behaved.  
In fact, it will be shown in  Section~\ref{sec:  ProdForm} that, subject 
to certain not very restrictive conditions on the operators $\hat{A}$ 
and $\hat{B}$, the sum on the right-hand side of  
Eq.~(\ref{eq:  HusProd}) is not only defined and asymptotic; it is
even absolutely convergent  for all $x$, $p$.  This is essentially
because $\hus{A}$ is typically a much less singular object than
$\weyl{A}$  [due to the Gaussian convolution in
Eq.~(\ref{eq:  HusDef})]. 

In Section~\ref{sec:  Expect} we apply the result just 
described to the problem of expressing expectation values in terms
of the Husimi function.

The expectation value of an operator $\hat{A}$ can be obtained
from the Wigner function   using the
formula~\cite{Hill,Lee}
\begin{equation}
  \Tr(\hat{\rho} \hat{A} )
= \int dx dp \, \weyl{A}(x,p) W(x,p)
\label{eq:  WigExpect}
\end{equation}
In certain cases we can  also express the expectation value in 
terms of the Husimi function  using~\cite{Hill,Lee,Miz}
\begin{equation}
 \Tr (\hat{\rho} \hat{A})
=  \int dx dp \, \anti{A}(x,p) \, \hfunc (x,p)
\label{eq:  HusExpect}
\end{equation}
where $\anti{A}$ is the anti-Husimi transform (or contravariant
symbol) of $\hat{A}$, defined by
\begin{equation}
  \anti{A}
=  e^{-\dplus \dminus} \hus{A}
\label{eq:  antiDef}
\end{equation}
[with $ \dplmin =2^{-1/2}(\partial_{x} \mp i \partial_{p})$, as
before].   Eq.~(\ref{eq:  HusExpect}) is valid (for example)
whenever~\cite{Reed} $\anti{A}$ exists as a tempered distribution
and
$\hfunc$ belongs to the corresponding space of test functions
(\emph{i.e.}\ the $C^{\infty}$ functions of rapid 
decrease). 
However, we have the problem that
$\anti{A}$ is often so highly singular that it is not defined as a
tempered distribution---which means that the usefulness of
Eq.~(\ref{eq:  HusExpect}) is somewhat limited.  This is often seen
as a serious drawback of the Husimi formalism.  

However, it turns out that it is often possible to circumvent this
difficulty.  Suppose we substitute the series given by 
Eq.~(\ref{eq:  antiDef}) into the right hand side of 
Eq.~(\ref{eq:  HusExpect}), and suppose we then reverse the 
order of sum and integral.  This gives
\begin{equation}
  \Tr (\hat{\rho} \hat{A})
= \sum_{n=0}^{\infty} \frac{(-1)^n}{n!}
  \int dx dp \, \left(\dplus^n \dminus^n \hus{A}(x,p) \right)
                  \hfunc(x,p)
\label{eq:  HusExpectB}
\end{equation}
In Section~\ref{sec:  Expect} we show that it often
happens that the sum on the right-hand side of this equation is
absolutely convergent, even in many of the cases where $\anti{A}$
fails to exist as a tempered distribution.

\section{Convergence of the Product Formula}
\label{sec:  ProdForm}
We will find it convenient to work in terms of coherent
states.  Define
\begin{equation*}
  \hat{a} = \frac{1}{\sqrt{2}}(\hat{x}+i \hat{p})
  \hspace{0.5 in}
  \hat{a}^{\dagger} = \frac{1}{\sqrt{2}}(\hat{x}-i \hat{p})
\end{equation*}
and let $\phi_{n}$ denote the $n^{\rm th}$
(normalised) eigenstate of the number operator
$\hat{a}^{\dagger} \hat{a}$:
\begin{equation*}
  \hat{a} \, \phi_{0} = 0 \hspace{0.5 in} 
  \phi_{n} = \frac{1}{\sqrt{n!}} (\hat{a}^{\dagger})^{ n}
\phi_{0}
\end{equation*}
Let $\hat{D}_{xp}$ be the displacement operator
\begin{equation*}
\hat{D}_{xp} = e^{i (p \hat{x} - x \hat{p})}
\end{equation*}
and define
\begin{equation*}
  \phi_{n; xp} = \hat{D}_{xp} \phi_{n}
  \hspace{0.5 in}
  \phi_{xp} = \phi_{0; xp} 
\end{equation*}
The $\phi_{xp}$ are the coherent states.  Let $\hat{A}$ be any
operator (not necessarily bounded) with domain of definition
$\mathscr{D}_{\hat{A}}$, and suppose that 
$\phi_{xp}\in \mathscr{D}_{\hat{A}}$ for all $x$, $p$. It is
then straightforward to show that
\begin{equation*}
  \hus{A}(x,p) = \vmat{\phi_{xp}}{\hat{A} \phi_{xp}}
\end{equation*}
(we no longer use the Dirac bra-ket notation, because
the existence of $\vmat{\psi}{ \hat{A} \chi}$ does not, in
general, imply the existence of
$\vmat{\hat{A}^{\dagger}\psi}{\chi}$).

If $\hat{A}$, $\hat{B}$ are both bounded then the proof of 
Eq.~(\ref{eq:  HusProd}) is comparatively straightforward. 
However, we want to make the proof as general as possible.  We
then have the difficulty that the sum in 
Eq.~(\ref{eq:  HusProd}) will only be defined 
if $\hus{A}$ and $\hus{B}$ are both $C^{\infty}$; whereas
functions of the form
$\vmat{\hat{D}_{xp}\psi}{ \hat{A}
\hat{D}_{xp} \psi}$ are, in general, not even
once-differentiable, let alone $C^{\infty}$.
We are thus faced with the question: what conditions
must we impose on the operator $\hat{A}$ in order 
to ensure that the function $\hus{A}$ is $C^{\infty}$?  One
answer to this question is given by the following theorem.
\begin{theorem}
\label{th:  a}
Let $\mathscr{D}_{\hat{A}}$,
$\mathscr{D}_{\hat{A}^{\dagger}}$ be the domains of definition 
of
$\hat{A}$, $\hat{A}^{\dagger}$ respectively.  Suppose that
$\phi_{xp} \in \mathscr{D}_{\hat{A}}\cap
\mathscr{D}_{\hat{A}^{\dagger}}$ for all $x$, $p$.  Suppose,
also, that $\vmat{\phi_{x_1 p_1}}{\hat{A} \phi_{x_2 p_2}}$,
$\vmat{\phi_{1; x_1 p_1}}{\hat{A} \phi_{x_2 p_2}}$
and
$\vmat{\phi_{1; x_1 p_1}}{\hat{A}^{\dagger} \phi_{x_2 p_2}}$
are continuous functions on $\mathbb{R}^4$.
Then $\hus{A}$ is an analytic function, which uniquely continues
to  a holomorphic function defined on the whole of
$\mathbb{C}^2$.

The continuation is given by
\begin{equation}
\hus{A}(x,p) 
= \frac{ \vmat{ \phi_{x_{-}p_{-}} }{
\hat{A}\phi_{x_{+}p_{+}}}
      }{\vmat{ \phi_{x_{-}p_{-}} }{ \phi_{x_{+}p_{+}} } }
\label{eq:  Continue}
\end{equation}
where $x,p$ are arbitrary complex, and where $x_{\pm}, p_{\pm}$
are the real variables defined by
\begin{align}
  x_{\pm} & = \frac{1}{2}(x+x^{*})\pm\frac{i}{2}(p-p^{*}) 
\label{eq:  xPmDef}
\\
  p_{\pm} & = \frac{1}{2}(p+p^{*})\mp\frac{i}{2}(x-x^{*})
\label{eq:  pPmDef}
\end{align}

\end{theorem}
This theorem is a strengthened version of results proved by
Mehta and Sudarshan~\cite{Mehta} and Appleby~\cite{self2}.
The proof  is given in 
Appendix~\ref{app:  ProofA}.

It is worth noting that the condition in the statement of this
theorem is quite weak.  If  the
three functions listed exist and are continuous then, without
making any  explicit assumption regarding the
differentiability of these functions, it automatically follows
that
$\hus{A}$ must be complex analytic.

We also have the following lemma:
\begin{lemma}
\label{lem:  b}
Suppose that $\hat{A}$ satisfies the conditions
of Theorem~\ref{th:  a}.  Then
\begin{align}
 \frac{  \vmat{\phi_{n;x_{-}p_{-}}
             }{\hat{A}\phi_{x_{+}p_{+}}}
      }{ \vmat{\phi_{x_{-}p_{-}}
             }{\phi_{x_{+}p_{+}} }
       }
& = \frac{1}{\sqrt{n!}}
     \sum_{r=0}^{n}  \begin{pmatrix} n \\ r\end{pmatrix}
          (z_{+}^{\vphantom{*}}-z_{-}^{*})^{n-r}
          \frac{\partial^r}{\partial z_{-}^{r} }
    \hus{A}(x,p)
\label{eq:  dBydZMinus}
\\
 \frac{  \vmat{\hat{A}^{\dagger}\phi_{x_{-}p_{-}}
             }{\phi_{n;x_{+}p_{+}}}
      }{ \vmat{\phi_{x_{-}p_{-}}
             }{\phi_{x_{+}p_{+}} }
       }
& = \frac{1}{\sqrt{n!}}
     \sum_{r=0}^{n}  \begin{pmatrix} n \\ r\end{pmatrix}
          (z_{-}^{\vphantom{*}}-z_{+}^{*})^{n-r}
          \frac{\partial^r}{\partial z_{+}^{r} }
    \hus{A}(x,p)
\label{eq:  dBydZPlus}
\end{align}
where $x_{\pm}$, $p_{\pm}$ are the variables defined 
by Eqs.~(\ref{eq:  xPmDef})
and~(\ref{eq:  pPmDef}), and where
\begin{equation*}
  z_{\pm} = \frac{1}{\sqrt{2}} (x \pm i p)
\end{equation*}
\end{lemma}
The proof of this lemma is given in  
Appendix~\ref{app:  ProofB}.

If $x$, $p$ are both real (so that $z_{-}=z_{+}^{*}$)
Eqs.~(\ref{eq:  dBydZMinus})
and~(\ref{eq:  dBydZPlus}) become
\begin{align}
  \vmat{ \phi_{n;x p} }{ \hat{A}\phi_{x p} }
& = \frac{1}{\sqrt{n!}} \, \dminus^{n} \hus{A}(x,p)
\label{eq:  dmnA}
\\
  \vmat{ \hat{A}^{\dagger}\phi_{x p} }{ \phi_{n;x p} }
& = \frac{1}{\sqrt{n!}} \, \dplus^{n} \hus{A}(x,p)
\label{eq:  dplA}
\end{align}
where
\begin{equation*}
  \dplmin = \frac{\partial}{\partial z_{\pm}}
  = \frac{1}{\sqrt{2}} 
  \left(\frac{\partial}{\partial x}
         \mp i \frac{\partial}{\partial p}
  \right)
\end{equation*}

Using these results the proof of the product formula becomes
very straightforward.  Let $\hat{A}$, $\hat{B}$ be any pair of 
operators satisfying the conditions of 
Theorem~\ref{th:  a}.  Suppose, also, that 
$\phi_{xp}\in \mathscr{D}_{\hat{A}\hat{B}}$ for all
$x, p \in \mathbb{R}$.  Then, for real $x$, $p$,
\allowdisplaybreaks{
\begin{align}
  \hus{(\hat{A}\hat{B})}(x,p)
& = \vmat{\phi_{xp}}{\hat{A}\hat{B}\phi_{xp}}
\notag
\\
& = \vmat{\hat{A}^{\dagger} \phi_{xp}}{\hat{B}\phi_{xp}}
\notag
\\
& = \sum_{n=0}^{\infty}
     \vmat{\hat{A}^{\dagger} \phi_{xp}}{\phi_{n; xp}}
     \vmat{\phi_{n; xp}}{\hat{B}\phi_{xp}}
\label{eq:  ProdInter}
\end{align}
where we have used the fact that the 
$\phi_{n; xp}$ constitute an orthonormal basis.  The absolute
convergence of this sum is an immediate consequence of basic
Hilbert space theory.  Using Eqs.~(\ref{eq:  dmnA})
and~(\ref{eq:  dplA}) we deduce
\begin{equation*}
  \hus{(\hat{A}\hat{B})}(x,p)
 = \sum_{n=0}^{\infty}
      \frac{1}{n!}
      \dplus^{n} \hus{A}(x,p) \dminus^{n} \hus{B}(x,p)
 = \hus{A}(x,p) \, 
      e^{\overleftarrow{\dplus}
         \overrightarrow{\dminus}}
     \hus{B}(x,p)
\end{equation*}
which is the product formula.  The absolute convergence of this
sum follows from the absolute convergence of the sum in
Eq.~(\ref{eq:  ProdInter}). }
\section{Expectation Values}
\label{sec:  Expect}
We now discuss the implications that the result just proved has
for the convergence of Eq.~(\ref{eq:  HusExpectB}), giving the
expectation value of $\hat{A}$ in terms of the Husimi function.

Of course, one does not expect  the right hand side of 
Eq.~(\ref{eq:  HusExpectB}) to converge for arbitrary $\hat{A}$
and $\hat{\rho}$---since, apart from anything else, an unbounded
operator does not have a well-defined expectation value for
every state
$\hat{\rho}$.  We therefore need to  place some kind of
restriction on the class of operators $\hat{A}$ and density
matrices $\hat{\rho}$ considered.  
The result we  prove is probably not the most general
possible.  However, it will serve to illustrate the point, that the
sum on the right hand side of Eq.~(\ref{eq:  HusExpectB}) is
often absolutely convergent, even in many of the cases where
the anti-Husimi transform fails to exist as a tempered
distribution.

We accordingly confine ourselves to the case of density
matrices for which the Husimi function 
$\hfunc \in \mathscr{I}(\mathbb{R}^2)$,
where $\mathscr{I}(\mathbb{R}^2)$ is the
space of $C^{\infty}$ functions which are rapidly decreasing at
infinity~\cite{Reed} (\emph{i.e.}\ the space of test functions
for the space of tempered distributions).  In other words, we
assume that
\begin{equation*}
 \sup_{(x,p)\in\mathbb{R}^2}
 \left|  (1+x^2+p^2)^l \partial_{x}^{m} \partial_{p}^{n} \hfunc
(x,p)
 \right|
< \infty
\end{equation*}
for every triplet of non-negative integers $l$, $m$, $n$.

We  assume that $\hat{A}$ has the properties
\begin{enumerate}
 \item $\phi_{xp} \in \mathscr{D}_{\hat{A}}
   \cap \mathscr{D}_{\hat{A}^{\dagger}}$ for all
   $x$,$p\in\mathbb{R}$.
 \item  There exist  positive constants $K_{\pm}$ and 
non-negative
  integers $N_{\pm}$ such that
  \begin{align}
    \| \hat{A}\phi_{xp}\| & \le K_{-} (1+x^2+p^2)^{N_{-}} 
    \label{eq:  Abound1}
    \\
    \| \hat{A}^{\dagger}\phi_{xp}\| 
       & \le K_{+} (1+x^2+p^2)^{N_{+}}
     \label{eq:  Abound2}
  \end{align}
  for all $x$, $p \in \mathbb{R}$.
\end{enumerate}
We will say that an operator satisfying these two conditions is
\emph{polynomial bounded}.  The following lemma gives two
 properties of such operators which will be needed in the
sequel.
\begin{lemma}
\label{lem:  c}
Suppose that $\hat{A}$ is polynomial bounded.  Then
\begin{enumerate}
\item $\hat{A}$ satisfies the conditions of 
Theorem~\ref{th:  a}.  In particular, $\hus{A}$ is analytic.
\item  For every pair of non-negative integers $m$, $n$ there
exists a positive constant $K_{mn}$ and a non-negative integer
$N_{mn}$ such that
\begin{equation}
   \left|\partial_{x}^{m} \partial_{p}^{n}\hus{A}(x,p)\right|
\le K_{mn} (1+x^2+p^2)^{N_{mn}}
\label{eq:  OMbound}
\end{equation}
for all $x$, $p\in\mathbb{R}$.
\end{enumerate}
\end{lemma}
The proof is given in 
Appendix~\ref{app:  ProofC}.

We now ready to prove the main result of this section.
\begin{theorem}
Suppose that $\hat{A}$ is polynomially bounded, and suppose
that the density matrix $\hat{\rho}$ is such that the
corresponding Husimi function is rapidly decreasing at
infinity.  Suppose, also, that $\hat{A}\hat{\rho}$ is of
trace-class.  Then 
\begin{equation}
  \Tr (\hat{A}\hat{\rho})
= \sum_{n=0}^{\infty} \frac{(-1)^n}{n!}
  \int dx dp \, \left(\dplus^n \dminus^n \hus{A}(x,p) \right)
                  \hfunc(x,p)
\label{eq:  HusExpectC}
\end{equation}
where the sum on the right hand side is absolutely convergent.
\end{theorem}
\begin{proof} We have
\begin{align*}
  \Tr(\hat{A}\hat{\rho})
& = \frac{1}{2 \pi} \int dx dp \,
         \vmat{\phi_{xp}}{\hat{A} \hat{\rho} \phi_{xp}}
\\
& =  \frac{1}{2 \pi} \int dx dp \, 
        \left( \sum_{n=0}^{\infty}
                \vmat{\hat{A}^{\dagger} \phi_{xp}}{
                      \phi_{n; xp}}
                 \vmat{\phi_{n; xp}}{\hat{\rho} \phi_{xp}}
        \right)
\end{align*}
We now use Lebesgue's dominated convergence
theorem~\cite{Reed} to show that we may reverse the order of
sum and integral.  In fact, it follows from
the Schwartz inequality that
\begin{align*}
&  \left| \sum_{n=0}^{m}
                \vmat{\hat{A}^{\dagger} \phi_{xp}}{
                      \phi_{n; xp}}
                 \vmat{\phi_{n; xp}}{\hat{\rho} \phi_{xp}} 
  \right|
\\
 & \hspace{1.0 in}
\le 
  \left( \biggl(\sum_{n=0}^{m} 
            \bigl|\vmat{\hat{A}^{\dagger} \phi_{xp}}{
                      \phi_{n; xp}}
            \bigr|^{2}
            \biggr)\biggl(
            \sum_{n=0}^{m} 
            \bigl|\vmat{\phi_{n; xp}}{\hat{\rho} \phi_{xp}}
            \bigr|^{2}
            \biggr)
  \right)^{\frac{1}{2}}
\\
&  \hspace{1.0 in}
\le 
\|\hat{A}^{\dagger} \phi_{xp}\| \;
\|\hat{\rho}\phi_{xp}\|
\end{align*}
We have
\begin{equation*}
  \| \hat{\rho}\phi_{xp} \|
= \bigl( \vmat{\phi_{xp}
                 }{\hat{\rho}^2 \phi_{xp}} 
    \bigr)^{\frac{1}{2}}
\le \bigl( \vmat{\phi_{xp}
                    }{\hat{\rho} \phi_{xp}} 
    \bigr)^{\frac{1}{2}}
= \sqrt{2 \pi Q(x,p)}
\end{equation*}
which, together with  Inequality~(\ref{eq:  Abound2}), implies
\begin{equation*}
\|\hat{A}^{\dagger} \phi_{xp}\| \;
\|\hat{\rho}\phi_{xp}\|
\le \sqrt{2\pi} K_{+}
       (1+x^2+p^2)^{N_{+}} \sqrt{Q(x,p)}
\end{equation*}
By assumption, $Q(x,p)\in\mathscr{I}(\mathbb{R}^2)$.  It
follows that $\|\hat{A}^{\dagger} \phi_{xp}\| \;
\|\hat{\rho}\phi_{xp}\|$ is integrable.  We may therefore use
Lebesgue's dominated convergence theorem~\cite{Reed} to
deduce
\begin{equation}
  \Tr(\hat{A}\hat{\rho})
 = 
 \frac{1}{2 \pi}\sum_{n=0}^{\infty}
  \int dx dp \, 
                \vmat{\hat{A}^{\dagger} \phi_{xp}}{
                      \phi_{n; xp}}
                 \vmat{\phi_{n; xp}}{\hat{\rho} \phi_{xp}}
\label{eq:  traceForm1}
\end{equation}
where the sum  is
absolutely convergent, since
\begin{equation*}
 \sum_{n=0}^{\infty}
  \left|
  \int dx dp \, 
                \vmat{\hat{A}^{\dagger} \phi_{xp}}{
                      \phi_{n; xp}}
                 \vmat{\phi_{n; xp}}{\hat{\rho} \phi_{xp}}
  \right|
\le
 \int dx dp \, 
\|\hat{A}^{\dagger} \phi_{xp}\| \;
\|\hat{\rho}\phi_{xp}\|
 < \infty
\end{equation*}
We know from Lemma~\ref{lem:  c} that $\hat{A}$ satisfies
the conditions of Theorem~\ref{th:  a}.  We may therefore use
the results proved in the last section to rewrite
Eq.~(\ref{eq:  traceForm1}) in the form
\begin{equation*}
  \Tr(\hat{A}\hat{\rho})
 = 
 \sum_{n=0}^{\infty}
  \frac{1}{n!}
  \int dx dp \, 
                \dplus^{n} \hus{A}(x,p)
                \dminus^{n} \hfunc (x,p)
\end{equation*}
Finally, it follows from Inequality~(\ref{eq:  OMbound}),
together with the fact that $Q\in \mathscr{I}(\mathbb{R}^2)$,
that we may partially integrate term-by-term to obtain
\begin{equation*}
  \Tr(\hat{A}\hat{\rho})
 = 
 \sum_{n=0}^{\infty}
  \frac{(-1)^{n}}{n!}
  \int dx dp \, 
              \bigl(
                \dplus^{n} \dminus^{n} \hus{A}(x,p)
               \bigr)
                 \hfunc (x,p)
\end{equation*}
\end{proof}

The right hand side of Eq.~(\ref{eq:  WigExpect}) (expressing
$\langle\hat{A}\rangle$ in terms of the Wigner function) is 
defined whenever $W\in \mathscr{I}(\mathbb{R}^2)$ and
$\weyl{A}$ exists as a tempered distribution.  On the other
hand, although it is true that
$\hfunc \in \mathscr{I}(\mathbb{R}^2)$ whenever
$W\in \mathscr{I}(\mathbb{R}^2)$ (see
Theorem~IX.3 of ref.~\cite{Reed}),  the fact that 
$\weyl{A}$ exists as a tempered distribution is not evidently
sufficient to ensure that $\hat{A}$ is polynomial bounded. 
So we have not  shown that 
Eq.~(\ref{eq:  HusExpectC}) has the same range of validity as
Eq.~(\ref{eq:  WigExpect}).    However, it can be shown 
that  $\hat{A}$ is polynomially bounded 
if $\phi_{xp}\in \mathscr{D}_{\hat{A}}\cap
\mathscr{D}_{\hat{A}^{\dagger}}$, and if
$\weyl{(\hat{A}^{\dagger} \hat{A})}$ and
$\weyl{(\hat{A}\hat{A}^{\dagger} )}$ exist as tempered
distributions  (see Theorem~IX.4 of ref.~\cite{Reed}).  In the
applications one meets with operators satisfying these conditions
much more commonly than one meets with operators for which
$\anti{A}$ exists as a tempered distribution.  For instance,
every bounded operator is polynomially bounded, whereas
there are many bounded operators of physical interest for which 
$\anti{A}$ fails to exist as a tempered distribution.  The above
result consequently represents a significant improvement on
the results that were previously known.

Of course, just from the fact that Eq.~(\ref{eq:  HusExpectC})
is convergent, it does not necessarily follow that the
convergence is sufficiently rapid to make the formula useful in
practical, numerical work.  This  question requires further
investigation.
\section{Conclusion}
\label{sec:  conc}
As has been stressed by Mizrahi~\cite{Miz},
Lalovi\'{c} \emph{et al}~\cite{Dav1},
Davidovi\'{c} and Lalovi\'{c}~\cite{Dav2} and others,
the Husimi formalism provides an especially perspicuous method
for studying the relationship between quantum and classical
mechanics. It establishes a one-to-one correspondence
between the basic equations of the two theories, so that one can
start with a classical formula, and then  turn it into the
corresponding quantum formula by adding successive correction
terms.  Moreover, the 
fact that $\hfunc(x,p)$
describes the outcome of a retrodictively optimal joint
measurement of $x$ and
$p$~\cite{Ali,self1}, means that one could reasonably argue that
the Husimi function is the most natural choice for a quantum
mechanical analogue of the classical probability distribution.

In this paper we have investigated the convergence properties
of two of the key formulae in the Husimi formalism.  We have
shown that the formula giving the Husimi transform of an
operator product has much better convergence properties than the
corresponding formula in the Wigner function formalism.
In particular, the Husimi formalism leads to  a
convergent generalization of the Liouville equation for a very
large class of Hamiltonians.  We have also shown that the
convergence properties of  the formula expressing
the expectation value
$\langle\hat{A}\rangle$ in terms of the Husimi function,
although seemingly not as good as those of the corresponding
formula in the Wigner function formalism,  are significantly
better than the often highly singular character of $\anti{A}$
would suggest.

These results lend  additional support to the
suggestion
that, in so far as the aim is specifically
to formulate quantum mechanics as a kind of generalized version
of classical mechanics, then the formalism based on the Husimi
function has some significant advantages.
\appendix
\section{Proof of Theorem~\ref{th:  a}}
\label{app:  ProofA}
For arbitrary complex $x$, $p$ define
\begin{equation*}
  F(x,p)
 = \frac{\vmat{\phi_{x_{-}p_{-}}}{\hat{A}\,\phi_{x_{+}p_{+}}}
       }{\vmat{\phi_{x_{-}p_{-}}}{\phi_{x_{+}p_{+}}}
       }
\end{equation*}
where $x_{\pm}$, $p_{\pm}$ are the (real) variables defined by
Eqs.~(\ref{eq:  xPmDef})
and~(\ref{eq:  pPmDef}).

It is easily seen that, if $x$, $p$ are both real, then
\begin{equation*}
 F(x,p) = \vmat{\phi_{x p}}{\hat{A}\, \phi_{x p}}
   = \hus{A}(x,p)
\end{equation*}
The problem thus reduces to that of showing that, if $\hat{A}$
has the properties stipulated,
then
$F$ is holomorphic.  We will do this by showing that
$F$ satisfies  the Cauchy-Riemann equations with respect to the
complex variables
\begin{equation}
  z_{\pm} = \frac{1}{\sqrt{2}} (x_{\pm}\pm i p_{\pm})
          = \frac{1}{\sqrt{2}} (x \pm i p)
\label{eq:  zPmDef}
\end{equation}

In fact, it is straightforward to show that $\phi_{xp}$,  
regarded as a vector-valued function of two real variables,  is
differentiable in the norm topology; the derivatives being
given by
\begin{align*}
 \frac{\partial}{\partial x} \phi_{xp}
& = \frac{1}{\sqrt{2}} \phi_{1; \, xp} -\frac{i}{2} p\phi_{xp}
\\
 \frac{\partial}{\partial p} \phi_{xp}
& = \frac{i}{\sqrt{2}} \phi_{1; \, xp} +\frac{i}{2} x\phi_{xp}
\end{align*}
Also,
\begin{equation*}
 \vmat{\phi_{x_{-}p_{-}}}{\phi_{x_{+}p_{+}}}
=  \exp\bigl[-\frac{1}{4}(x_{+}-x_{-})^2  
              -\frac{1}{4}(p_{+}-p_{-})^2
              +\frac{i}{2}(p_{+}x_{-}-p_{-}x_{+})\bigr]
\end{equation*}
Consequently, $F$ is differentiable with respect to the 
variables $x_{-}$, $p_{-}$.  Moreover
\begin{align}
\frac{\partial}{\partial x_{-}} F(x,p)
& = \frac{1}{\sqrt{2}} 
  \frac{\vmat{\phi_{1;\,x_{-}p_{-}}
            }{\hat{A}\phi_{x_{+}p_{+}}}
       }{\vmat{\phi_{x_{-}p_{-}}}{\phi_{x_{+}p_{+}}}}
   +\frac{1}{\sqrt{2}} (z_{-}^{*}-z_{+})F(x,p)
\notag
\\
& = 
i \frac{\partial}{\partial p_{-}} F(x,p)
\label{eq:  CauRieA}
\end{align}
from which it follows that $F$ satisfies the Cauchy-Riemann
equations with respect to the complex variable 
$z_{-}=(x_{-}-ip_{-})/\sqrt{2}$.

We can alternatively write
\begin{equation*}
  F(x,p)
 =
\frac{\vmat{\hat{A}^{\dagger}\, \phi_{x_{-}p_{-}}
          }{\phi_{x_{+}p_{+}}}
     }{\vmat{\phi_{x_{-}p_{-}}}{\phi_{x_{+}p_{+}}}
       }
\end{equation*}
Consequently, $F$ is also differentiable with respect to the
real variables $x_{+}$, $p_{+}$.  Moreover
\begin{align}
\frac{\partial}{\partial x_{+}} F(x,p)
& = \frac{1}{\sqrt{2}} 
  \frac{\vmat{\hat{A}^{\dagger}\, \phi_{x_{-}p_{-}}
            }{\phi_{1;\, x_{+}p_{+}}}
       }{\vmat{\phi_{x_{-}p_{-}}}{\phi_{x_{+}p_{+}}}}
   +\frac{1}{\sqrt{2}} (z_{+}^{*}-z_{-})F(x,p)
\notag
\\
& = 
-i \frac{\partial}{\partial p_{+}} F(x,p)
\label{eq:  CauRieB}
\end{align}
from which it follows that $F$ satisfies the Cauchy-Riemann
equations with respect to the complex variable 
$z_{+}=(x_{+}+ip_{+})/\sqrt{2}$.

If $\hat{A}$ has the properties specified in the statement of
theorem, then we see from  Eqs.~(\ref{eq:  CauRieA})
and~(\ref{eq:  CauRieB}) that the partial derivatives
$\partial F/\partial x_{\pm}$, 
$\partial F/\partial p_{\pm}$, are continuous functions on
$\mathbb{R}^4$.  It follows~\cite{Grau} that $F$ is a
holomorphic function of the complex variables $z_{\pm}$.
Referring to Eq.~(\ref{eq:  zPmDef}) it can be seen that
the variables $z_{\pm}$ are linear combinations of  $x$, $p$.
We conclude that $F$ is  a holomorphic function of  $x$,
$p$.
\section{Proof of Lemma~\ref{lem:  b}}
\label{app:  ProofB}
It is straightforward to show that 
$\phi_{n; x_{-} p_{-}}$, regarded
as a vector valued function of two real variables, is
differentiable in the norm topology.  Moreover
\begin{equation*}
  \frac{1}{\sqrt{2}}
  \left( \frac{\partial}{\partial x_{-}}
        -i \frac{\partial}{\partial p_{-}}
  \right) \phi_{n ; x_{-} p_{-}}
 = \sqrt{n+1} \phi_{(n+1); x_{-} p_{-}}
 +\frac{1}{2} z_{-}\, \phi_{n; x_{-} p_{-}}
\end{equation*}
Hence
\begin{multline*}
  \frac{1}{\sqrt{2}}
  \left( \frac{\partial}{\partial x_{-}}
        +i \frac{\partial}{\partial p_{-}}
  \right)
  \frac{  \vmat{\phi_{n;x_{-}p_{-}}
             }{\hat{A}\phi_{x_{+}p_{+}}}
      }{ \vmat{\phi_{x_{-}p_{-}}
             }{\phi_{x_{+}p_{+}} }
       }
\\
= \sqrt{n+1}
  \frac{  \vmat{\phi_{(n+1);x_{-}p_{-}}
             }{\hat{A}\phi_{x_{+}p_{+}}}
      }{ \vmat{\phi_{x_{-}p_{-}}
             }{\phi_{x_{+}p_{+}} }
       }
       + \left(z_{-}^{*}-z_{+}
          \right)
    \frac{  \vmat{\phi_{n;x_{-}p_{-}}
             }{\hat{A}\phi_{x_{+}p_{+}}}
      }{ \vmat{\phi_{x_{-}p_{-}}
             }{\phi_{x_{+}p_{+}} }
       }
\end{multline*}
Iterating this result, and using
\begin{equation*}
  \frac{1}{2^{r/2}}
  \left(\frac{\partial}{\partial x_{-}}+
          i\frac{\partial}{\partial p_{-}}
  \right)^{r}
  \hus{A}(x,p)
= \frac{\partial^r}{\partial z_{-}^{r}}
  \hus{A}(x,p)
\end{equation*}
we obtain
Eq.~(\ref{eq:  dBydZMinus}).  

The proof of Eq.~(\ref{eq:  dBydZPlus}) is
similar.
\section{Proof of Lemma~\ref{lem:  c}}
\label{app:  ProofC}
\subsection*{Proof of (1)}  We need to show that, if $\hat{A}$
is polynomial bounded, then the functions
$\vmat{\phi_{x_1 p_1}}{\hat{A} \phi_{x_2 p_2}}$,
$\vmat{\phi_{1; x_1 p_1}}{\hat{A} \phi_{x_2 p_2}}$
and
$\vmat{\phi_{1; x_1 p_1}}{\hat{A}^{\dagger} \phi_{x_2 p_2}}$
are continuous.

Consider the function $\vmat{\phi_{1; x_1 p_1}}{\hat{A}
\phi_{x_2 p_2}}$. We have
\begin{multline*}
\left|\bvmat{\phi_{1; {x'\vphantom{p}}_{\!\! 1} {p'}_{\!\! 1}}
           }{\hat{A} \phi_{{x'\vphantom{p}}_{\!\! 2} 
           {p'}_{\!\! 2} }}
     -\bvmat{\phi_{1; x_1 p_1}}{\hat{A} \phi_{x_2 p_2}}
\right|
\\
\le
\left|\bvmat{ (\phi_{1; {x'\vphantom{p}}_{\!\! 1} 
                           {p'}_{\!\! 1}}
               -\phi_{1; x_1 p_1})
          }{\hat{A} \phi_{{x'\vphantom{p}}_{\!\! 2} 
                    {p'}_{\!\! 2}} }
\right| +
\left|\bvmat{ \phi_{1; x_1 p_1}
          }{\hat{A}( \phi_{{x'\vphantom{p}}_{\!\! 2} 
                    {p'}_{\!\! 2}}- \phi_{x_2 p_2})}
\right|
\end{multline*}
In view of Eq.~(\ref{eq:  Abound1}) we have
\begin{equation*}
  \left|\bvmat{ (\phi_{1; {x'\vphantom{p}}_{\!\! 1} 
                    {p'}_{\!\! 1}}-\phi_{1;x_1 p_1})
          }{\hat{A} \phi_{{x'\vphantom{p}}_{\!\! 2} 
                            {p'}_{\!\! 2}}}
  \right|
\le 
K_{-} (1+{x'\vphantom{p}}_{\!\! 2}^{2}+
                 {p'}_{\!\! 2}^{2})^{N_{-}}
\|\phi_{1; {x'\vphantom{p}}_{\!\! 1} 
         {p'}_{\!\! 1}}-\phi_{1; x_1 p_1} \|
\end{equation*}
Also, using the completeness relation for coherent states, together with
Eq.~(\ref{eq:  Abound2}), we find
\begin{align*}
& \left|\bvmat{ \phi_{1; x_1 p_1}
          }{\hat{A}( \phi_{{x' \vphantom{p}}_{\!\! 2} 
                  {p'}_{\!\! 2}}-
             \phi_{x_2 p_2})}
\right|
\\
& \hspace{0.8 in}
=
\left|\frac{1}{2 \pi} \int dx_{3} dp_{3} \,
    \bvmat{ \phi_{1; x_1 p_1}
          }{\phi_{x_3 p_3}}
    \bvmat{\hat{A}^{\dagger}\phi_{x_3 p_3}
         }{( \phi_{{x'\vphantom{p}}_{\!\! 2} 
               {p'}_{\!\! 2}}-
\phi_{x_2 p_2})}
\right|
\\
& \hspace{0.8 in}
\le f(x_1,p_1) \| \phi_{{x'\vphantom{p}}_{\!\! 2} 
                   {p'}_{\!\! 2}}-
\phi_{x_2 p_2} \|
\end{align*}
where $f$ is the polynomial
\begin{align*}
  f(x_1,p_1)
& = \frac{K_{+}}{2 \pi}
     \int dx_3 dp_3 \,
          \bigl|\bvmat{ \phi_{1; x_1 p_1}
                     }{\phi_{x_3 p_3}}
          \bigr|
          (1+x_{3}^{2}+p_{3}^{2})^{N_{+}}
\\
& = \frac{K_{+}}{2^{\frac{3}{2}} \pi}
     \int d{x'\vphantom{p}}_{\! \! 3} 
                    d{p'}_{\! \! 3} \,
       \sqrt{1+{x'\vphantom{p}}_{\! \!  3}^{2} 
       + {p'}_{\! \! 3}^{2}}
       \left(1+({x'\vphantom{p}}_{\! \! 3}+
       x_{1})^2+({p'}_{\! \! 3}+p_{1})^2
       \right)^{N_{+}}
\\
& \hspace{2.5 in} \times
       \exp\left[-\frac{1}{4}
       \bigl({x'\vphantom{p}}_{\! \! 3}^{2}
                  + {p'}_{\! \!3}^{2}\bigr)
           \right]
\end{align*}
Putting these results together we find
\begin{multline*}
\left|\bvmat{\phi_{1; {x'\vphantom{p}}_{\!\! 1} {p'}_{\!\! 1}}
           }{\hat{A} \phi_{{x'\vphantom{p}}_{\!\! 2} 
           {p'}_{\!\! 2} }}
     -\bvmat{\phi_{1; x_1 p_1}}{\hat{A} \phi_{x_2 p_2}}
\right|
\\
\le
K_{-} (1+{x'\vphantom{p}}_{\!\! 2}^{2}+
                 {p'}_{\!\! 2}^{2})^{N_{-}}
\|\phi_{1; {x'\vphantom{p}}_{\!\! 1} 
         {p'}_{\!\! 1}}-\phi_{1; x_1 p_1} \|
+
f(x_1,p_1) \| \phi_{{x'\vphantom{p}}_{\!\! 2} 
                   {p'}_{\!\! 2}}-
\phi_{x_2 p_2} \|
\end{multline*}
$\phi_{x p}$ and $\phi_{1; x p}$, regarded as
vector-valued functions on $\mathbb{R}^2$, are continuous in
the norm topology.  Consequently
\begin{equation*}
\bvmat{\phi_{1; {x'\vphantom{p}}_{\!\! 1} {p'}_{\!\! 1} }
           }{
           \hat{A} \phi_{{x'\vphantom{p}}_{\!\! 2}
             {p'}_{\!\! 1}}
           }
\rightarrow
\bvmat{\phi_{1; x_1 p_1}}{\hat{A} \phi_{x_2 p_2}}
\end{equation*}
as $({x'\vphantom{p}}_{\!\! 1}, {p'}_{\!\! 1},
{x'\vphantom{p}}_{\!\! 2}, {p'}_{\!\! 2})
\rightarrow (x^{\vphantom{j}}_1 , p^{\vphantom{j}}_1,
x^{\vphantom{j}}_2 , p^{\vphantom{j}}_2)$.  It follows that 
$\bvmat{\phi_{1; x_1 p_1}}{\hat{A} \phi_{x_2 p_2}}$ is
continuous.  Continuity of the functions
$\bvmat{\phi_{x_1 p_1}}{\hat{A} \phi_{x_2 p_2}}$
and
$\bvmat{\phi_{1; x_1 p_1}}{\hat{A}^{\dagger}
\phi_{x_2 p_2}}$ is proved in the same way.
\subsection*{Proof of (2).}  The completeness relation for coherent states
implies
\begin{equation*}
  \hus{A}(x,p) = \vmat{\phi_{xp}}{\hat{A}\phi_{xp}}
=  \frac{1}{2 \pi} \int dx' dp' \,
   \vmat{\phi_{xp}}{\phi_{x' p'}}
   \vmat{\hat{A}^{\dagger}\phi_{x' p'}}{\phi_{xp}}
\end{equation*}
Using
\begin{align*}
       \dplus \phi_{n; xp} 
& = \sqrt{n+1}\phi_{(n+1);xp} + \frac{1}{2}z^{*}\phi_{n;xp}
\\
      \dminus \phi_{n; xp}
& =
     \begin{cases}
          -\frac{1}{2} z \phi_{xp} 
                       \hspace{0.5 in} & \text{if}\; n=0 \\
          -\sqrt{n} \phi_{(n-1); xp}
            -\frac{1}{2} z \phi_{n; xp}
              \hspace{0.5 in} & \text{if}\; n>0
     \end{cases}
\end{align*}
[where $\dplmin = 2^{-1/2}(\partial_{x}\mp i \partial_{p})$
and $z=2^{-1/2}(x+i p)$)], and differentiating under the integral
sign,  it is not difficult to show that
\begin{equation*}
\partial_{x}^{m} \partial_{p}^{n} \hus{A}(x,p)
=
\sum_{r,s=0}^{n+m}
  c_{r s}
   \int dx' dp' \,
   \vmat{\phi_{r; xp}}{\phi_{x' p'}}
   \vmat{\hat{A}^{\dagger}\phi_{x' p'}}{\phi_{s; xp}}
\end{equation*}
for suitable constants $c_{r s}$.
We have
\begin{equation*}
  \bigl|   \vmat{\phi_{r; xp}}{\phi_{x' p'}}
  \bigr|
=  \frac{1}{2^{\frac{r}{2}}\sqrt{r!}}
    \left( (x'-x)^2 +(p'-p)^2
    \right)^{\frac{r}{2}}
    \exp\left[ -\frac{1}{4}(x'-x)^2 -\frac{1}{4}(p'-p)^2
    \right]
\end{equation*}
In view of Inequality~(\ref{eq:  Abound2}) it follows
that
 \begin{multline*}
\left|   \int dx' dp' \,
   \vmat{\phi_{r; xp}}{\phi_{x' p'}}
   \vmat{\hat{A}^{\dagger}\phi_{x' p'}}{\phi_{s; xp}}
  \right|
\\  \le 
  \frac{K_{+}}{2^{\frac{r}{2}}\sqrt{r!}}
    \int dx'' dp'' \,
        \bigl( 1 + (x''+x)^2 +(p''+p)^2
        \bigr)^{N_{+}}
\\
    \times
      \bigl({x''}^{2} +{p''}^{2}\bigr)^{\frac{r}{2}}
      \exp \left[ -\frac{1}{4} \bigl({x''}^{2}+{p''}^{2}\bigr)
      \right]
\end{multline*}
It can be seen that the expression on the right hand side of this
inequality is a polynomial in $x$ and $p$.  Consequently
\begin{equation*}
\left| \partial_{x}^{m} \partial_{p}^{n} \hus{A}(x,p)
\right|
\le
f(x,p)
\end{equation*}
for some polynomial $f(x,p)$.  The claim is now immediate.

\end{document}